\documentclass[12pt,epsf]{article}
\usepackage{graphics}
\input epsf

\begin{document}
\baselineskip 24pt

\newcommand{\beq}{\begin{equation}}
\newcommand{\eeq}{\end{equation}}
\newcommand{\bea}{\begin{eqnarray}}
\newcommand{\eea}{\end{eqnarray}}

\def \lsim{\mathrel{\vcenter
     {\hbox{$<$}\nointerlineskip\hbox{$\sim$}}}}
\def \gsim{\mathrel{\vcenter
     {\hbox{$>$}\nointerlineskip\hbox{$\sim$}}}}

\newcommand{\sheptitle}
{Testing the see-saw mechanism at collider energies}

\newcommand{\shepauthor}
{S. F. King$^1$ and T. Yanagida$^2$}

\newcommand{\shepaddress}
{$^1$School of Physics and Astronomy, University of Southampton, \\
        Southampton, SO17 1BJ, U.K\\
\vspace{0.25in}
$^2$Department of Physics, University of Tokyo,\\
Tokyo 113-0033, Japan}

\vspace{0.25in}

\newcommand{\shepabstract}
{We propose a low energy extension of the Standard Model
consisting of an additional gauged $U(1)_{B-L}$ plus three
right-handed neutrinos.
The lightest right-handed neutrinos have TeV scale masses
and may be produced at colliders via their couplings to the
$Z_{B-L}$ gauge boson whose mass and gauge coupling is 
constrained by the out-of-equilibrium condition
leading to upper bounds on the right-handed neutrino 
and $Z_{B-L}$ production cross-sections at colliders.
We propose a brane-world scenario which motivates
such TeV mass right-handed neutrinos.
Our analysis opens up the possibility that the
mechanism responsible for neutrino mass
is testable at colliders such as the LHC or VLHC.}

\begin{titlepage}
\begin{flushright}
hep-ph/0411030 \\
\end{flushright}
\begin{center}
{\large{\bf \sheptitle}}
\\ \shepauthor \\ \mbox{} \\ {\it \shepaddress} \\
{\bf Abstract} \bigskip \end{center} \setcounter{page}{0}
\shepabstract
\begin{flushleft}
\today
\end{flushleft}

\vskip 0.1in
\noindent

\end{titlepage}

\newpage

\section{Introduction}
The see-saw mechanism \cite{seesaw1},\cite{seesaw2}
is an attractive mechanism for accounting
for light neutrino masses. The mechanism works by introducing
right-handed neutrinos with ``large'' Majorana masses, which violate
lepton number $L$, and the Yukawa couplings between the left-handed
leptons and right-handed neutrinos then results in small effective 
Majorana mass operators for left-handed neutrinos.
Such a scenario has potentially important cosmological implications 
for the baryon asymmetry of the universe via a mechanism known as 
leptogenesis \cite{yanagida1}. It is clear that the see-saw mechanism
satisfies the Sakharov conditions of lepton number violation, 
and CP violation due to the 
complex neutrino Yukawa couplings.
Providing the out-of-equilibrium condition
is also met when the right-handed neutrinos
decay, a net lepton number $L$ may then be generated which may 
subsequently be converted into 
a net baryon number $B$ by sphaleron interactions which preserve $B-L$. 

Given the recent progress in neutrino physics, there has been much
discussion concerning both the see-saw mechanism 
\cite{King:2003jb} and leptogenesis \cite{Giudice:2003jh}.
In the simplest implementations of thermal leptogenesis
the re-heat temperature after inflation must be in excess
of the lower bound arising from gravitino production
\cite{Kawasaki:2004qu}.
\footnote{In the type II see-saw mechanism this conflict can be 
somewhat ameliorated \cite{Antusch:2004xy}.}
Typically thermal leptogenesis works best when the lightest
right-handed neutrino mass exceeds about $10^9$ GeV, while in order
to avoid excessive thermal production of gravitinos the temperature
of the universe after inflation must not exceed this value. 
Even ignoring the gravitino problem, such large values of
right-handed neutrino masses make them inaccessible to 
experiment at planned or even imagined collider energies.
Testing the see-saw mechanism experimentally in any direct way
therefore seems virtually impossible under the thermal leptogenesis
framework.

A possible solution to the gravitino problem is provided by the
idea of resonant leptogenesis \cite{Pilaftsis:2003gt}.
The key observation of resonant leptogenesis 
is that, if the lightest two right-handed neutrinos
are closely degenerate, then resonance effects can enhance the
production of lepton number even for a right-handed neutrino mass
scale as low as a TeV, allowing the reheat temperature 
to be low enough to avoid the gravitino problem.
\footnote{If supersymmetry (SUSY) is additionally assumed then TeV mass
right-handed neutrinos allow the possibility that the soft SUSY
breaking mass parameters may be at least partly responsible
for leptogenesis and radiatively generated neutrino masses
\cite{Boubekeur:2004ez}.
The present analysis can be extended for this case.}

With such light right-handed neutrinos one may think it possible
to test the see-saw mechanism
experimentally at collider energies. However in the absence of 
additional interactions this is not the case since,
even if TeV mass right-handed neutrinos
are kinematically accessible to high energy colliders such as
the LHC, their Yukawa couplings are necessarily so weak 
as to render their production cross-section unobservable.

In this paper we propose a minimal scenario in which, in addition
to having light right-handed neutrinos, there is also
a low energy gauged $B-L$ symmetry $U(1)_{B-L}$.
\footnote{In general the extra $U(1)$ could be some linear combination
of $B-L$ and hypercharge $Y$. The vector space spanned by these
generators includes the third generator of $SU(2)_R$, $T_{3R}$,
and the $U(1)_X$ generator $X$ contained in the maximal $SO(10)$ subgroup
$SU(5)\times U(1)_X$. But here we focus on the case of an
extra $U(1)_{B-L}$ for simplicity and definiteness.}
It is natural that the scale of gauged $B-L$ symmetry breaking should be
somewhat higher than the heaviest right-handed neutrino mass, 
since gauged $B-L$ symmetry forbids Majorana neutrino masses.
Cancellation of gauged $B-L$ anomalies requires that there should be three
right-handed neutrinos.
If we assume the lightest pair of right-handed
neutrinos to be degenerate, we would therefore expect that the
third right-handed neutrino mass $M_3$
to be closer to the mass scale
of $B-L$ symmetry breaking $v_{B-L}$, which is somewhat higher than the
lightest right-handed neutrinos mass scale $M_1$.
We are therefore led to propose a new low energy extension of the Standard
Model consisting of an additional gauged $U(1)_{B-L}$ plus three
right-handed neutrinos with the following new mass scales:
$M_1\lsim M_2 \lsim M_3 \lsim v_{B-L}$, with a possible accurate degeneracy
involving the lightest right-handed neutrino pair.

The main goal of this paper is to 
discuss the collider phenomenology of 
light right-handed neutrinos and $Z_{B-L}$ gauge bosons,
constrained by the out-of-equilibrium conditions
as required by thermal leptogenesis.
\footnote{Note that above its breaking scale $v_{B-L}$ the presence 
of exact gauged $B-L$ means that the net $B-L$ of the
universe must be exactly zero. Since $B+L$ is violated by sphaleron
interactions this implies that baryogenesis or leptogenesis cannot
occur above the scale $v_{B-L}$.}
We shall require the effective leptogenesis neutrino
mass parameter to be $\tilde{m}_1\sim 10^{-3}$ eV
for efficient production and out-of-equibrium decay
of the right-handed neutrinos $N_{1,2}$.
\footnote{Our scenario should not be confused with a recently 
discussed scenario in which low energy leptogenesis occurs due to 
scattering from domain walls produced by the breaking 
of a discrete left-right symmetry,
and the right-handed neutrinos are constrained
not to erase the $B-L$ \cite{Sahu:2004sb}.}
The requirement that the 
lightest right-handed neutrino pair be out-of-equilibrium when they
decay implies that the $B-L$ interactions must be sufficiently
weak, leading to a lower bound on $v_{B-L}$
or an upper bound on the gauge coupling $g_{B-L}$, depending on the
mass ordering of $M_1$ and $M_{B-L}$.
These bounds in turn leads to upper limits on the
production cross-section for right-handed neutrinos and $Z_{B-L}$
gauge bosons at colliders, which we shall also discuss.
We shall also explore the theoretical motivation for 
TeV mass right-handed neutrinos, and propose a specific brane-world
scenario.

\section{Out-of-equilibrium Condition}

\subsection{$M_{B-L}\gg M_1$}

We shall begin by assuming that $M_{B-L}\gg M_1$, and show that 
the out-of-equilibrium condition leads to a lower bound on
$v_{B-L}$ in this case.

The lightest right-handed neutrinos must decay while they are
out-of-equilibrium in the early Universe.
We have already assumed that the direct decays of right-handed neutrinos 
are always out-of-equilibrium, consistent with
$\tilde{m}_1\sim 10^{-3}$ eV, so that we only need additionally
ensure that the new $Z_{B-L}$ interactions do not
bring the lightest right-handed neutrinos back into thermal equilibrium.
\footnote{Note that the new $Z_{B-L}$ interactions do not violate
lepton number by themselves. However since the right-handed neutrinos are
Majorana particles, such $Z_{B-L}$ interactions 
(with unsuppressed L-violating mass insertions) 
would bring the right-handed neutrinos in the thermal equibrium with
vanishing chemical potential unless they are out of equilibrium.
}
The reaction rate of right-handed neutrinos is given by:
\beq
\Gamma =<\sigma_{ann} n v>
\label{gamma}
\eeq
where $\sigma_{ann}$ is the total annihilation cross-section
of lightest right-handed neutrinos into three families of Standard
Model fermions ($f$) and antifermions ($\bar{f}$),
and $n$ is the number density of right-handed neutrinos.
\footnote{Note that the annihilation cross-section is the most relevant
one for satisfying the out-of-equilibrium condition. 
We have also assumed that scalar fermions are heavier than
the right-handed neutrinos $N_1$ and $N_2$.}
The annihilation cross-section $\sigma_{ann}$ is given by:
\beq
\sigma_{ann}=\sigma (N_1 N_1 \rightarrow \sum_f f\bar{f})
\eeq
due to the tree-level exchange of a $B-L$ gauge boson 
of mass $M_{B-L}$ with a gauge coupling $g_{B-L}$,
\beq
\sigma (N_1 N_1 \rightarrow \sum_f f\bar{f}) \sim 
3\times \frac{13}{3}\frac{g_{B-L}^4}{48\pi}\frac{E^2}{M_{B-L}^4}
\sim \frac{1}{4\pi}\frac{E^2}{v_{B-L}^4}
\label{sigma}
\eeq
where the mass of the gauge boson is given by $M_{B-L}=g_{B-L}v_{B-L}$.
The number density of right-handed neutrinos $n$ given by:
\beq
n=\frac{3}{4}\frac{2.404}{2\pi^2}g\left(\frac{kT}{\hbar c}\right)^3
\sim \frac{1}{\pi^2}T^3
\label{n}
\eeq
setting $\hbar = c=k=1$. 

To generate lepton number asymmetry, the right-handed neutrinos must be
out-of-equilibrium when they decay. If they decay at a temperature 
$E\sim T \sim M_1$ then their reaction rate when they decay is given
from Eqs.(\ref{gamma},\ref{sigma},\ref{n}) by
\beq
\Gamma =<\sigma_{ann} n v> \sim
\frac{1}{4\pi}\frac{E^2}{v_{B-L}^4}.\frac{1}{\pi^2}T^3 \sim 
\frac{1}{4\pi^3}\frac{M_1^5}{v_{B-L}^4}. 
\label{gamma2}
\eeq
To be out-of-equilibrium the reaction rate be less than the
Hubble constant $H$ whose square at a temperature $T$ is given by
\beq
H^2\approx \frac{8\pi}{3}G_N\rho 
\sim \frac{4\pi^3}{45}\frac{g^*T^4}{M_P^2}\sim 3\frac{g^*T^4}{M_P^2}
\label{H}
\eeq
setting $\hbar = c=k=1$ where $g^*$ is the total number of degrees of
freedom at the temperature $T$ and $M_P$ is the Planck mass
$M_P\approx 1.2\times 10^{19}$ GeV.
The out-of-equilibrium condition 
is given by $\Gamma < aH$ at $E\sim T \sim M_1$, where $a\sim O(10)$,
which from
Eqs.(\ref{gamma2},\ref{H}) leads to the lower bound on $v_{B-L}$:
\beq
v_{B-L} > \left( \frac{M_P}{4\pi^3a\sqrt{3g^*}}\right)^{1/4}M_1^{3/4}
\sim 10^6\ {\rm GeV} \left( \frac{M_1}{1\ {\rm TeV}}  \right)^{3/4}.
\label{bound}
\eeq
Eq.(\ref{bound}) tells us that for a degenerate
pair of right-handed neutrinos of mass $1\ {\rm TeV}$ the scale of
$B-L$ breaking must exceed $10^6\ {\rm GeV}$. 
\footnote{ A similar constraint was considered
in left-right symmetric models \cite{Mohapatra:1992pk}.}

\subsection{$M_{B-L}\lsim 2M_1$}

We now consider the case that $M_{B-L}\lsim 2M_1$, and show that 
the out-of-equilibrium condition leads to an upper bound on
$g_{B-L}$ in this case.

In this case the estimate in Eq.(\ref{sigma}) becomes
\beq
\sigma (N_1 N_1 \rightarrow \sum_f f\bar{f}) \sim 
3\times \frac{13}{3}\frac{g_{B-L}^4}{48\pi}\frac{1}{16E^2}
\sim \frac{1}{4\pi}\frac{g_{B-L}^4}{16E^2}.
\label{sigma2}
\eeq
The resulting reaction rate at $E\sim T \sim M_1$ in Eq.(\ref{gamma2}) 
becomes modified to
\beq
\Gamma =<\sigma_{ann} n v> \sim
\frac{1}{4\pi^3}\left(\frac{g_{B-L}}{2}\right)^4 M_1. 
\label{gamma3}
\eeq
The out-of-equilibrium condition 
is given by $\Gamma < aH$ at $E\sim T \sim M_1$, where $a\sim O(10)$,
which from
Eqs.(\ref{gamma3},\ref{H}) leads to the upper bound on $g_{B-L}$:
\beq
g_{B-L} < 2\times \left[ 4\pi^3 a \sqrt{3g^*}\left( \frac{M_1}{M_P}\right)\right]^{1/4}
\sim 2\times 10^{-3}\left( \frac{M_1}{1\ {\rm TeV}}  \right)^{1/4}.
\label{bound2}
\eeq

Other reactions are important at $E\sim T \sim M_1$ for the case
$M_{B-L}< 2M_1$, for example real $Z_{B-L}$ pair production
and annihilation: $N_1N_1\leftrightarrow  Z_{B-L}Z_{B-L}$ 
whose cross-section will be of similar magnitude to that in Eq.(\ref{sigma2}).

For $M_{B-L} \sim 2M_1$ at $E\sim T \sim M_1$ it becomes possible to 
have single $Z_{B-L}$ production and decay:
$ Z_{B-L} \leftrightarrow f\bar{f} $ (including $N_1N_1$).
This special case leads to a
quite different bound on the gauge coupling since the decay rate
of $Z_{B-L}$ is given by:
\beq
\Gamma ( Z_{B-L} \rightarrow \sum_f f\bar{f} )
\sim 3\times \frac{13}{3}\frac{g_{B-L}^2}{48\pi}M_1 
\sim \frac{1}{4\pi}g_{B-L}^2M_1. 
\label{gamma4}
\eeq
The out-of-equilibrium condition in this case is obtained by comparing
the decay rate in Eq.(\ref{gamma4}) to the Hubble expansion rate
$\Gamma < aH$ at $E\sim T \sim M_1$, where $a\sim O(10)$,
which from
Eqs.(\ref{gamma4},\ref{H}) leads to the upper bound on $g_{B-L}$:
\beq
g_{B-L} < \sqrt{4\pi a}(3g^*)^{1/4}\left( \frac{M_1}{M_P}  \right)^{1/2}
\sim 10^{-6}\left( \frac{M_1}{1\ {\rm TeV}}  \right)^{1/2}.
\label{bound3}
\eeq
In this special case when $M_{B-L} \sim 2M_1$ the bound on the gauge
coupling in Eq.(\ref{bound3}) is much more stringent than the 
bound on the gauge coupling in Eq.(\ref{bound2}) for the more general
case $M_{B-L}< 2M_1$.

Finally we should note that the upper bound on $g_{B-L}$ is obtained even for
the case, $M_{B-L} \lsim 10M_1$. For $M_{B-L}> M_1$ the inverse decay
has a Boltzmann suppression $e^{M_{B-L}/T}$ with $T\simeq M_1$. 
The Boltzmann factor is only $10^{-3}$ for $M_{B-L} \simeq 10M_1$ and hence 
we may have still a strigent bound on $g_{B-L}^2$ 
but weaker than the case $M_{B-L} \sim 2M_1$. So the bound on $v_{B-L}$
is given when $M_{B-L} \gsim 20\times M_1$ 
since the Boltzmann factor is $10^{-6}$ and
the inverse decay of $Z_{B-L}$ becomes negligible compared with the $B-L$ gauge
exchanges.

\section{Phenomenology}

For the case of a heavy $Z_{B-L}$, 
$M_{B-L}\gg M_1$, the typical cross-section for 
production of the lightest right-handed neutrinos at colliders is
given from Eq.(\ref{sigma}) and bounded from Eq.(\ref{bound}) by
\beq
\sigma (e^+e^-\rightarrow N_1 N_1) <
\frac{1}{48\pi}\frac{E^2}{10^{24}\ {\rm GeV^4}}
\left( \frac{1\ {\rm TeV}}{M_1}  \right)^{3}
\sim 
{3\times 10^{-9}\ {\rm fb}}\left( \frac{E}{1\ {\rm TeV}} \right)^{2}
\left( \frac{1\ {\rm TeV}}{M_1}\right)^{3}
\label{sigma3}
\eeq
where we have used the conversion 
$1\ {\rm GeV}^{-2}\approx 4\times 10^{-4}b$.
Unfortunately the cross-section may be too small to enable
right-handed neutrinos to be produced at TeV energies such as the LHC
or CLIC. For efficient production of right-handed neutrinos a collider
with an energy approaching the mass of the $B-L$ gauge boson
$M_{B-L}=g_{B-L}v_{B-L}$ would be required, such as the VLHC
for example. Although the right-handed neutrinos are produced
in pairs via their $Z_{B-L}$ couplings, they will decay singly
via their small Yukawa couplings into either (left-handed) 
neutrino plus Higgs $h^0$, or charged lepton plus (longitudinal) $W$,
with a characteristic signature in both cases.
The expected decay rate for TeV mass right-handed neutrinos 
with a Yukawa coupling about $10^{-6}$ would be 
$\Gamma(N_1\rightarrow HL)\approx 10^{-10}$ GeV corresponding to a
lifetime of about $10^{-14}$s.

Turning to the other possibility of a lighter $Z_{B-L}$, with
$M_{B-L}< 2M_1$, from Eq.(\ref{bound2}) we see that the gauge coupling
must be smaller than about $10^{-3}$ for TeV scale right-handed
neutrinos. This also leads to a cross-section for
right-handed neutrinos at colliders bounded by
\beq
\sigma (e^+e^-\rightarrow N_1 N_1) <
{4\times 10^{-7}\ {\rm fb}}\left( \frac{1\ {\rm TeV}}{E} \right)^{2}
\left( \frac{M_1}{1\ {\rm TeV}}\right).
\label{sigma4}
\eeq
Unlike the cross-section in Eq.(\ref{sigma3}), the cross-section in 
Eq.(\ref{sigma4}) is suppressed at higher energies. However, in this
case one might hope to produce real on-shell $Z_{B-L}$
with mass below the TeV scale. The $Z_{B-L}$ may be produced
singly by mechanisms which involve only two powers of $g_{B-L}$
in the rate, and so the typical production cross-sections are expected to 
be enhanced relative to that in Eq.(\ref{sigma4}) by a factor of about
a million. For example
the $Z_{B-L}$ may be produced singly by the Drell-Yan
process, and may be discovered via its decays into $e^+,e^-$,
$\mu^+,\mu^-$, or jet-jet. The discovery limits for such a
$Z'$ at the LHC have been studied by ATLAS, and for example a TeV mass
$Z'$ may be discovered at the 5$\sigma$ level with 100$fb^{-1}$ 
providing the square of the gauge coupling is about $10^{-3}$
\cite{atlas}.
With the small gauge coupling bounded in Eq.(\ref{bound2}), discovery at the
LHC will be clearly difficult, but since this bound
is independent of the mass of the $Z_{B-L}$, which may be as light as
the ordinary $Z$ for example, 
its discovery may be possible in principle at the LHC.
Note that 
the couplings of the $B-L$ gauge boson $Z_{B-L}$ to leptons and quarks
are 1:1/3, which may be tested if its decay is observed. 

\section{A Brane Scenario}

In this section we address the following questions:
\begin{enumerate}
\item Why are the Yukawa couplings inducing the Dirac neutrino masses
    so small ? To get neutrino masses of order $10^{-2}$ eV with 1 TeV
    right-handed Majorana mass we need $M_{Dirac} \simeq 10^{-4}$ GeV.
    Thus the Yukawa coupling should be $\sim 10^{-6}$.
\item Is there a natural inflation model with reheat temperature 
    $T_R \simeq$ a few TeV ?
    In other words, why do we assume $M_1\simeq$ 1 TeV in the
    estimates above? The constraint
    from the gravitino problem tells us only $T_R < 10^{6}$ GeV and hence
    $M_1$ could be as large as $10^{6}$ GeV.
\end{enumerate}

The answer to the second question is straighforward:
lower-energy scale inflation may produce lower reheating temperature, 
in general. The new inflation model is a candidate for the low-energy scale
inflation. One of us has 
constructed a new inflation model in SUGRA and found that
the reheating temperature is given by $T_R\simeq m_{3/2} <$ few TeV
\cite{Yanagida}.

Moreover, the low-energy scale inflation is
favored in string landscape. The string theory may have a number of vacuua
which may have many inflaton candidates. The vacuum which has many inflations
is favored, since many inflations make larger universe. So it is better to have
as lower-energy scale inflation as possible and it is likely that a new
inflation
is the last inflation we can see.
\footnote{T.Y. thanks K.-I. Izawa for the discussion of inflation models on the string landscape.}

The answer to the first question is not difficult also.
Small right-handed neutrino masses at the TeV scale which require
small Dirac Yukawa couplings in the range $10^{-5}-10^{-6}$.
To account for this we propose the following extra-dimensional set up
consisting of two parallel 3-branes where the standard model
fermions and Higgs are on one brane, and the right-handed neutrinos
are on the other brane, and the $B-L$ gauge multiplet is in the bulk which
contains $n$ extra dimensions. 
The $B-L$ gauge interaction has 4 dimensional anomaly on the each
branes, but it is cancelled by bulk Chern-Simons term 
\cite{Yanagida2}.
The $B-L$ is broken by a Higgs vacuum 
expectation value $v_{B-L}$ located on one of the 3-branes.
The Yukawa coupling constant in this case is exponentially suppressed
$Y_{\nu}\sim e^{-M_*L}$ where $M_*$ is the cut-off (string) scale
and $L$ is the compactification scale of the extra dimension(s).
Assuming $M_*L \sim 11-14$ can account for 
Dirac Yukawa couplings in the range $10^{-5}-10^{-6}$.

In such a scenario the gauge coupling of the $B-L$ 
is given by 
\beq
g_{B-L}\sim g_0(M_*L)^{-n/2}
\eeq
Assuming $g_0\sim 0.3$ and $M_*L \sim 12$ we find
$g_{B-L}\sim 2\times 10^{-2}$ for $n=2$ 
which implies the $B-L$ gauge boson mass 
$M_{B-L} \sim 20$ TeV for $v_{B-L} \simeq 10^6$ GeV. 
For n=6 we find $g_{B-L}\sim 2\times 10^{-4}$,
which is in the appropriate range for the light
$Z_{B-L}$ scenario

\section{Conclusion}
We have proposed a low energy extension of the Standard Model
consisting of an additional gauged $U(1)_{B-L}$ plus three
right-handed neutrinos where the lightest right-handed
neutrino mass scale is $M_1\gsim 1$ TeV. 
In the absence of SUSY, this requires resonant leptogenesis
with the lightest right-handed neutrino pair being 
approximately degenerate. 

We have discussed the collider phenomenology of 
light right-handed neutrinos and $Z_{B-L}$ gauge bosons,
constrained by the out-of-equilibrium conditions.
We find that for $M_{B-L}\gg M_1$ there is 
a lower limit on the symmetry breaking scale $v_{B-L}$, while for 
$M_{B-L}\lsim 2M_1$ there is an upper limit on
the gauge coupling $g_{B-L}$. Although the 
TeV mass right-handed neutrinos may be 
produced at colliders via their couplings to the
$Z_{B-L}$ gauge bosons, the above limits severely constain
the production cross-sections of both right-handed neutrinos 
and $Z_{B-L}$ gauge bosons at colliders.

We have also considered the
theoretical motivation for TeV scale right-handed neutrinos
coming from brane-world set-ups and string theory.
We have proposed a particular brane-world scenario in which small
Yukawa couplings emerge, with a $B-L$ gauge coupling 
and mass in the appropriate ranges consistent with the bounds
from thermal leptogenesis, depending on the number of extra
dimensions. For two extra dimensions the mass of the 
$Z_{B-L}$ gauge boson is expected to have a mass $M_{B-L}\sim 20$ TeV.
The cross-section for production of right-handed neutrinos
is expected to become observable in this case when
the centre of mass energy of the collider approaches
$M_{B-L}$, which motivates a future collider such as the VLHC. 
For six extra dimensions the $Z_{B-L}$ gauge boson could be as light as the
ordinary $Z$ boson with a cross-sections for production
that will make its discovery at the LHC challenging.

In conclusion, the possibility of TeV mass right-handed
neutrinos, together with additional $Z_{B-L}$, is cosmologically
consistent from the point of view of leptogenesis,
and has some theoretical motivation from string theory and extra
dimensions. It would open up the
possibility of testing the mechanism responsible for
neutrino mass experimentally 
at collider energies corresponding to the LHC or a future VLHC.

\begin{center}
{\bf Acknowledgments}
\end{center}
We would like to thank the organisers of the Nobel Symposium 129
on Neutrino Physics where this work was begun.
S.K. thanks Stefan Antusch for useful comments.

\end{document}